\documentclass[11pt]{article}
\usepackage{color}
\usepackage{amscd}
\usepackage[psamsfonts]{amssymb}
\usepackage{beig,amssymb,amsmath}
\input{psfig}

\newtheorem{propos}{Proposition}[section]
\newtheorem{lem}{Lemma}[section]

\newtheorem{defin}{Definition}[section]

\newcounter{remarques}[section]

\renewcommand{\theremarques}{\arabic{remarques}}

\def\Rm{\refstepcounter{remarques}{\bf Remark \theremarques} }
\def\ali{\hfill\break}
\def\w{\omega}
\def\W{\Omega}
\def\unN{\{1\ldots ,N\}}

\def\demi{{1\over 2}}
\def\ddd{{\mathcal D}}

\def\hhh{{\mathcal H}}

\def\nn{{<n>}}
\def\ND{{ND}}
\def\infi{{<\infty>}}
\def\intX{{\stackrel{\circ}{X}}}
\def\intF{{\stackrel{\circ}{F}}}

\def\unN{{\{1,\ldots ,N\}}}

\def\j1p0{{j_1^0,\ldots ,j_p^0}}

\def\jp1{{j_p,\ldots ,j_1}}
\def\j1p{{j_1,\ldots ,j_p}}

\def\BR{{\Bbb R}}
\def\supp{{\hbox{supp}}}

\def\BN{{\Bbb N}}
\def\BZ{{\Bbb Z}}
\def\BQ{{\Bbb Q}}

\def\BP{{\Bbb P}}
\def\BE{{\Bbb E}}

\def\beq{\begin{eqnarray}}
\def\eeq{\end{eqnarray}}
\def\beqn{\begin{eqnarray*}}
\def\eeqn{\end{eqnarray*}}
\def\bpr{\begin{propos}}
\def\epr{\end{propos}}
\def\indic{{1}}

\def\11{{(1,1)}}

\def\Qu01{{Q^{u_0,u_1}}}

\begin{document}
\centerline{\Large \bf
Schr\"odinger operators on fractal lattices
}
\centerline{\Large\bf
with random blow-ups}
\ali
\ali
\centerline{\Large\bf Christophe Sabot}
\ali
\begin{center}
Ecole Normale Sup\'erieure,
\\
DMA, 45, rue d'Ulm, 75005, Paris,
\\
and,
\\
Laboratoire de Probabilit\'es et mod\`eles
al\'eatoires, Universit\'e Paris 6,
\\
4, Place Jussieu, 75252 Paris cedex 5
\footnote{E-mail adress:
sabot@ccr.jussieu.fr}
\end{center}
\ali
\ali
\ali
{\bf Abstract:}
Starting from a finitely ramified self-similar set $X$
we can construct an unbounded set $X_\infi$
by blowing-up the initial set $X$.
We consider random blow-ups and prove elementary
properties of the spectrum of the
natural Laplace operator on $X_\infi$ (and on the associated lattice).
We prove that the spectral type of the operator is almost surely
deterministic with the blow-up and that the spectrum coincides
with the support of the density of states almost surely
(actually our result is more precise).
We also prove that if the density of states is completely created by
the so-called Neuman-Dirichlet eigenvalues, then almost surely
the spectrum is
pure point with compactly supported eigenfunctions.
\ali
\ali
\ali
\ali
\ali\ali\ali
\vfill\break

In this text we prove elementary results on spectral properties
of Laplace operators on unbounded fractal lattices based
on finitely ramified self-similar sets, and there
continuous analogous.
One of the main novelty in this text
is to consider random blow-up, i.e. the unbounded lattice or fractal
is constructed by blowing-up randomly the initial figure.
The results of this text show that the natural Laplace operator
on these random lattices share the same basic properties as  ergodic
families of random Schr\"odinger operators, as defined for
example in \cite{PasturF}, \cite{CarmonaL}.
In particular, we consider the relations between the spectrum of
the operator and two important measures: the density of
states and the density of Neuman-Dirichlet eigenvalues
(also called molecular states in \cite{Rammal}) which are
eigenvalues associated with eigenfunctions with both
Neuman and Dirichlet boundary condition.
In \cite{Sabot5} we investigated the relations between
these measures and the dynamics of a certain renormalization
map, which is a rational map of a compact K\"ahler manifold.

Starting from a finitely ramified self-similar set $X$
(for example a p.c.f self-similar set as in \cite{Kigami1})
we can construct an increasing sequence $X_\nn$ by blowing-up $X$.
The way the set $X$ is blowed-up is represented by
a sequence $\w=(\w_1,\ldots ,\w_n,\ldots )$
in $\unN^\BN$, where $N$ is the number of cells in $X$.
The unbounded set $X_\infi$ is defined as the union $X_\infi=\cup_n X_\nn$.
If $H=H_{<0>}$ is the "natural" Laplace operator on $X$ then we can
define by scaling a sequence of operators $H_\nn$ on $X_\nn$ 
and  $H_\infi$ on $X_\infi$.
The density of states (resp. of Neuman-Dirichlet eigenvalues)
is defined as the limit of the renormalized counting measures
of the eingenvalues of $H_\nn$ (resp. of the Neuman-Dirichlet eigenvalues of 
$H_\nn$).
In section 2 we prove three elementary results. The first two are  the 
counterpart of well-known properties of ergodic families of
random Schr\"odinger operators.
The third one is more specific to our situation since it
involves the Neuman-Dirichlet spectrum which is empty in the
case of Schr\"odinger operators on $\BZ^d$.
\begin{itemize}
\item
In proposition (\ref{p.1}) we prove that almost surely on
$\w$, the support of the density of states is equal
to the spectrum of the operator on $X_\infi$ 
(actually, we can precise for which $\w$ this equality is always true).
\item
In proposition (\ref{p.2}) we prove that the spectral type of the operator is 
almost surely deterministic, i.e.
that there exists deterministic subsets $\Sigma,\; \Sigma_{ac},\;
\Sigma_{sc},\;\Sigma_{pp}$  such that almost surely in $\w$,
$\Sigma,\; \Sigma_{ac},\;
\Sigma_{sc},\;\Sigma_{pp}$ are respectively the spectrum,
the absolutely continuous spectrum, the singular continuous spectrum
and the pure point spectrum of the operator on $X_\infi$.
\item
In proposition (\ref{p.3}) 
we prove that if the density of states is completely created
by the Neuman-Dirichlet eigenvalues (i.e. if the density
of states is equal to the density of Neuman-Dirichlet eigenvalues)
then the spectrum of the operator on $X_\infi$ is pure point with
compactly supported eigenfunctions, almost surely in $\w$.
This result is important since in \cite{Sabot5}, theorem 4.1 and
proposition 4.4,
we proved
that this happen exactly when the asymptotic degree of the 
renormalization map is smaller than $N$, the number of cells in
$X$.
\end{itemize}
In section 3 we introduce several measures that generalize the
density of states and the density of Neuman-Dirichlet eigenvalues
to the different parts of the spectrum.
The result of proposition \ref{p.2}
suggests that
the right object to investigate is the almost sure type of the spectrum
of the operator on $X_\infi$.
A step is done in this direction in \cite{Sabot5}, where the density of states
and the density of N-D eigenvalues are computed in terms of
a certain explicite renormalization map.
Previously, for some particular examples the 
spectral properties has been investigated, 
cf \cite{MalozemovT}, \cite{Teplyaev}, \cite{Sabot4}.
In particular in \cite{Teplyaev}, Teplyaev investigated 
the spectrum of the discrete Laplace operator
on the lattice associated with the Sierpinski gasket for
different blow-ups.

In the first section we briefly recall
the notations, but we send  to the main text \cite{Sabot5}
for precise definitions, examples and a more complete
bibliography.

\subsection{Notations}
\subsubsection{Self-similar sets and self-similar lattices.}
We first briefly recall the notations and definitions of \cite{Sabot5}.
Suppose that $X$ is a finitely ramified self-similar set
as defined in \cite{Sabot5} and, to simplify notations,
that $X$ has a geometrical embedding in $\BR^d$.
This means that $X$ is a proper compact connected subset of
$\BR^d$ and that there exists $N$ strictly contractive
similitudes $(\Psi_1,\ldots ,\Psi_N)$, with distinct fixed points, 
such that
$$X=\cup_{i=1}^N \Psi_i(X),$$
and that there exists a subset $F$ of the set of fixed points
$(x_1,\ldots ,x_N)$ of $(\Psi_1,\ldots ,\Psi_N)$ such that
$$\Psi_i(X)\cap \Psi_j(X)=\Psi_i(F)\cap \Psi_j(F),\;\;\;\forall i\ne j.$$

Set $\W=\unN^\BN$ and fix an element $\w$ of $\W$.
We define the blow-up of $X$ as the sequence of increasing
sets $X_\nn$ defined by $X_{<0>}=X$ and 
$$X_\nn=\Psi^{-1}_{\w_1}\circ \cdots \circ \Psi_{w_n}^{-1} (X).
$$
We set $X_\infi=\cup_{n=0}^\infty X_\nn$.
The boundary of $X_{<0>}=X$ is defined as $\partial X=F$ and we set
$$\partial X_\nn =\Psi^{-1}_{\w_1}\circ \cdots \circ \Psi_{w_n}^{-1} (F),$$
and $\partial X_\infi =\cap_n \cup_{m\ge n} \partial X_{<m>}$.
We also set $\intX_\nn= X_\nn\setminus \partial X_\nn$
(and similarly for $X_\infi$).
Remark that
$$X_{<n+p>}
=\cup_{j_1,\ldots ,j_p} X_{<n+p>,j_1,\ldots ,j_p},
$$ where
$$ X_{<n+p>,j_1,\ldots ,j_p}=
\Psi_{\w_1}^{-1}\circ \cdots \circ \Psi_{\w_{n+p}}^{-1}
(\Psi_{j_1}\circ \cdots \circ\Psi_{j_p}(X)).$$
We call the $X_{<n+p>,j_1,\ldots ,j_p}$ the $<n>$-cells of
$X_{<n+p>}$ and we remark that $$X_{<n>}=X_{<n+p>,\w_{n+p},\ldots,\w_{n+1}}$$.
Remark also that $X_{<n+p>,j_1,\ldots ,j_p}$ is naturally isomorphic to
$X_{<n>}$.

This structure of increasing sets has a discrete counterpart.
Set $F_{<0>}=F$ and 
$$F_{<n>}=\Psi^{-1}_{\w_1}\circ \cdots \Psi_{w_n}^{-1}(\cup_{j_1,\ldots ,j_n=1}^N
\Psi_{j_1}\circ\cdots\circ \Psi_{j_n}(F)).$$
(Hence, $F_{<n>}$ is the union of the boundaries of the 0-cells
$X_{\nn,j_1,\ldots ,j_n}$ of $X_\nn$.)
The sequence $F_\nn$ is clearly increasing and we set
$F_\infi=\cup_n F_\nn$.
Similarly, we have
$$F_{<n+p>}
=\cup_{j_1,\ldots ,j_p} F_{<n+p>,j_1,\ldots ,j_p},
$$ where $F_{<n+p>,j_1,\ldots ,j_p}$ are the $<n>$-cells of $F_{<n+p>}$
defined by
$$ F_{<n+p>,j_1,\ldots ,j_p}=
\Psi_{\w_1}^{-1}\circ \cdots \circ \Psi_{\w_{n+p}}^{-1}
\circ \Psi_{j_1}\circ \cdots \circ\Psi_{j_p}
(\cup_{i_1,\ldots, i_n} \Psi_{i_1}\circ \cdots \circ \Psi_{i_n}(F)).$$
We set $\partial F_\nn=\partial X_\nn$ and $\intF_\nn= F_\nn\setminus \partial F_\nn$
(and idem for $\partial F_\infi$).

\subsubsection{Self-similar Laplacians.}
We fix for the rest of the text two $N$-tuples
$(\alpha_1,\ldots ,\alpha_N)\in ]0,1[^N$
and $(\beta_1,\ldots ,\beta_N)\in ]0,1[^N$ such that 
$\beta_1+\cdots \beta_N=1$.
The $N$-tuple $(\alpha_1,\ldots ,\alpha_N)$, resp. $(\beta_1,\ldots ,\beta_N)$
will represent the scaling in energy, resp. in measure in our structure.
We set $\gamma_i=(\alpha_i \beta_i)^{-1}$ and
we make the following assumption
\ali

(H) We suppose that $(\beta_1,\ldots ,\beta_N)$ is
proportional to $(\alpha_1^{-1},\ldots ,\alpha_N^{-1})$
so that $\gamma_i$ does not depend on $i$. We denote by
$\gamma$ the common value of the $\gamma_i$.
\ali
\ali
{\bf
Construction in the discrete case.}

To construct a dicrete Laplace operator on the sequence of
lattices $F_\nn$ we suppose given $A$,  a non-negative
symmetric endomorphism  of
$\BR^F$ of the form
\beq
\label{f.1.5}
Af(x)=-\sum_{y\in F, y\ne x } a_{x,y} (f(y)-f(x)), \;\;\;\forall f\in \BR^F, 
\forall x\in F,
\eeq
where $a_{x,y}$, $x\ne y$, are non negative reals such that
$a_{x,y}=a_{y,x}$.
We suppose moreover that $A$ is irreducible, i.e. that the graph on $F$ defined
by strictly positive $a_{x,y}$ is connected.
We suppose also given a strictly positive measure $b$ on $F$.
\ali
We denote by $A_{<n>,i_1,\ldots ,i_n}$ 
(resp. $b_{\nn,i_1,\ldots ,i_n}$) the copy of the operator $A$ 
(resp. of the measure $b$) 
on the cell
$F_{\nn, i_1,\ldots ,i_n}$ (cf \cite{Sabot5} for precise definition).
Then we define the symmetric operator $A_\nn$ and the measure $b_\nn$ on
$F_\nn$ by
\beq
\label{f.1.6}
A_\nn =\sum_{i_1,\ldots ,i_n=1}^N 
\alpha_{\w_1}\cdots \alpha_{\w_n} \alpha^{-1}_{i_1}\cdots \alpha_{i_n}^{-1}
A_{\nn,i_1,\ldots ,i_n},
\\
\label{f.1.7}
b_\nn =\sum_{i_1,\ldots ,i_n=1}^N 
\beta_{\w_1}^{-1}\cdots \beta_{\w_n}^{-1} 
\beta_{i_1}\cdots \beta_{i_n}
b_{\nn,i_1,\ldots ,i_n}.
\eeq
\Rm:
\label{r.1}
We see from the definition that the value of
$A_\nn$ and $b_\nn$ depend on the $N$-tuples $(\alpha_1,\ldots ,\alpha_N)$
and $(\beta_1,\ldots ,\beta_N)$ only up to a constant.
\ali
\ali
Remark that $A_\nn$ and $b_\nn$ form an inductive sequence since if
$\supp (f)\subset \intF_{<p>}$ for $p\le n$ then
$$A_\nn f = A_{<p>} f \;\;\;\hbox{and}\;\;\; \int f db_\nn =\int f db_{<p>}.$$
Therefore $A_\nn$ and $b_\nn$ can be extended respectively
to a  linear operator $A_\infi$ on $\BR^{F_\infi}$ and to a measure $b_\infi$ on
$F_\infi$.
\ali
Remark that since $X$ is connected, $A_\nn$ is irreducible, i.e.
$A_\nn f=0$ implies that $f$ is a constant function.
Denote by $<\cdot ,\cdot >$ the usual scalar product on
$\BR^{F_\nn}$.
Let $H_\nn^+$ be the operator on $L^2(F_\nn, b_\nn)$ defined by:
\beq
\label{f.1.8}
<A_\nn f,g>=- \int H_\nn^+ f g db_\nn\;\;\; \forall f,g\in \BR^{F_\nn}.
\eeq
The operator $H_\nn^+$ is non-positive,
self-adjoint on $L^2(F_\nn, b_\nn)$.
The operator with Dirichlet boundary condition, denoted $H_\nn^-$, is the
self-adjoint operator on $\BR^{\intF_\nn}$ defined as the restriction
of $H_\nn^+$ to
$\BR^{\intF_\nn}\simeq \{f \in \BR^{F_\nn},\;\; f_{|\partial F_\nn}=0\}$.
To be coherent with the notations of the continuous case we
sometimes write $\ddd_\nn^+=\BR^{F_\nn}$ and $\ddd_\nn^-=\BR^{\intF_\nn}$
for the domains of $H_\nn^\pm$.

If $K>0$ is such that $<Af,f>\le K \int f^2 db$ for all $f$ in
$\BR^F$ then it is easy to see from (\ref{f.1.6}) and (\ref{f.1.7}) and
assumption (H)
that the same inequality is true for
$A_\nn$ and $b_\nn$ and for $A_\infi$ et $b_\infi$. 
Thus the sequence $H_\nn^\pm$ is uniformly bounded
for the operator norm on  $L^2(b_\nn)$ and can be extended into a
non-positive, self-adjoint operator $H_\infi^+$ on $\ddd_\infi^+=L^2(b_\infi)$.
We define $H^-_\infi$ as the restriction of $H_\infi^+$ to
$\ddd_\infi^-=\{f\in \ddd_\infi^+,\; f_{|\partial F_\infi}=0\}$.
Clearly, we have
$$<A_\infi f,g>=-\int H_\infi^\pm f g 
db_\infi,\;\;\; \forall f,g\in \ddd_\infi^\pm.
$$
Remark that if $\partial F_\infi =\emptyset $ then the operators $H_\infi^+$
and $H_\infi^-$ are equal and in this case we simply write $H_\infi$ for
$H_\infi^+=H_\infi^-$.

Note finally that the hypothesis (H) corresponds to a property of
local invariance by
translation of the operator $H_\infi^\pm$, as explained in \cite{Sabot5}
(this assumption is related to the lattice case condition introduced
in \cite{KigamiL}, \cite{Lapidus}).
\ali
\ali
{\bf In the continuous case}

We know that there exists a unique positive measure on $X$,
which is self-similar with respect to the weights $(\beta_1,\ldots ,\beta_N)$,
i.e. which statisfies
\beqn
\int_X fdm =\sum_{i=1}^N \beta_i \int_X f\circ \Psi_i dm.
\eeqn
We suppose given on $X$ a local, regular, conservative Dirichlet form
$(a,\ddd)$ on $L^2(X,m)$, self-similar with respect to the weights
$(\alpha_1,\ldots ,\alpha_N)$, as defined in \cite{Sabot1}.
Essentially, this means that $a$ satisfies
\beqn
a(f,f)=\sum_{i=1}^N (\alpha_i)^{-1} a(f\circ \Psi_i,f\circ \Psi_i ).
\eeqn

On $X_\nn$ we define the measures $m_\nn$ and the Dirichlet form
$(a_\nn,\ddd_\nn)$ by scaling by
$$ \int_{X_\nn}  f dm_\nn = \beta_{\w_1}^{-1}\cdots \beta_{\w_n}^{-1}
\int_X f\circ \Psi_{\w_1}^{-1} \circ \cdots \circ \Psi_{\w_n}^{-1} dm,
\;\;\;\forall f\in C^0(X_\nn),$$
and
\beqn
\ddd_\nn=\{f\in L^2(m_\nn),\hbox{ s.t. }  
f\circ \Psi_{\w_1}^{-1} \circ \cdots \circ \Psi_{\w_n}^{-1}\in \ddd\},
\\
a_\nn (f)=\alpha_{\w_1}\cdots \alpha_{\w_n} 
a(f\circ \Psi_{\w_1}^{-1} \circ \cdots \circ \Psi_{\w_n}^{-1}).
\eeqn
If $f$ in $\ddd_{<n+p>}$ is such that
$\supp (f)\subset \intX_\nn$ then we see that
$a_{<n+p>} (f,f)= a_\nn(f,f)$ and $\int f dm_\nn =\int fdm_{<n+p>}$.
Hence, we see that $m_{\nn}$ can be extended to a measure $m_{<\infty>}$ on
$X_\infi$, and we set
$$\ddd_\infi =\{f\in L^2(X_\infi, m_\infi),
\;\; \sup_n a_\nn (f_{|X_\nn}, f_{|X_\nn})<\infty\},
$$
and 
$a_\infi (f,f)=\lim_{n\to\infty} a_\nn (f_{|X_\nn}, f_{|X_\nn})$ on
$\ddd_\infi$.
We set $\ddd_\nn^-=\{f\in \ddd_\nn,\;\; f_{|\partial X_\nn} =0\}$
and $\ddd^+_\nn =\ddd_\nn$ (and idem for $\ddd^\pm_\infi$).
We define $H_\nn^\pm$ and $H^\pm_\infi$ as the infinitesimal generators of
$(a_\nn,\ddd_\nn^\pm)$ and $(a_\infi,\ddd_\infi^\pm)$.
\ali
We refer to \cite{Sabot5} for examples.

\subsubsection{The density of states and the density of Neuman-Dirichlet eigenvalues.}
We denote both in the continuous case and in the lattice
case by $\nu_\nn^\pm$ the counting measure of the eigenvalues of
the operators $H_\nn^\pm$ (in the lattice $\nu_\nn^\pm$ is a finite sum of
Dirac masses, in the continuous case it is a countable sums of Dirac masses
accumulating at infinity).
As usual, the density of states, that we denote $\mu$,
is defined as the limit (when it exists
and is the same for Neuman and Dirichlet boundary condition)
$$\mu=\lim_{n\to\infty} {1\over N^n } \nu_\nn^\pm.$$
The existence of this measure is proved
in \cite{Fukushima2}, \cite{KigamiL}.
(Remark that despite the terminology, $\mu$ is a measure which does
not necessarly have a density).
\ali

We say that a function f is a Neuman-Dirichlet (N-D for short) eigenfunction
of $H_{<n>}$ with eigenvalue $\lambda$ if it is both an eigenfunction
of $H_\nn^-$ and $H_\nn^+$, i.e.
in the lattice case this means that
$f$ is $\ddd_\nn^-$, i.e.
$f_{|\partial F_{\nn}}=0$, and that
$$<A_\nn f, g>=-\lambda \int f g db_{<n>},\;\;\;
\forall g\in \ddd_\nn^+=\BR^{F_{<n>}},$$
and in the continuous case that
$f$ is in $\ddd_{<n>}^-$,
and that
$$a_\nn (f, g)=-\lambda \int f g dm_{<n>}, \;\;\;
\forall g\in \ddd_{<n>}^+.$$
We denote by $\nu^\ND_\nn$ the counting measure of the N-D eigenvalues of $H_\nn$
(counted with multiplicity) and by $E_\nn^\ND$ the subspace of $\ddd_\nn^+$
generated by the N-D eigenfunctions.

Remark that any function $f$ of $E_\nn^\ND$, when extended by 0 to
$F_{<n+p>}$ (resp. $X_{<n+p>}$) is a N-D eigenfunction of $H_{<n+p>}$.
When extended by 0 to
$F_\infi$ (resp.  $X_\infi$)  it is an eigenfunction of
$H_\infi^+$ and $H_\infi^-$, with compact support.
We denote by $\hhh_\ND$ the closure in $\ddd_\infi^+$ of the space
$\cup_n E_{<n>}^\ND$.
\ali
It is easy to see that
$$\nu^\ND_{<n+1>}\ge N \nu_\nn^\ND.$$
Indeed, if $f$ is a N-D eigenfunction of $H_\nn$ then
we can construct $N$ copies of $f$ on the $N$ $\nn$-cells
of $F_{<n+1>}$. Precisely, for all $i=1,\ldots ,N$
we consider the function
$f_{i}$ on $\BR^{F_{<n+1>}}$ which is the copy of $f$ on
$F_{<n+1>,i}$ and equal to $0$ on $F_{<n+1>}\setminus F_{<n+1>,i}$.
These functions form an orthogonal familly of N-D eigenfunctions of
$H_{<n+1>}$ with same eigenvalues 
(by the hypothesis (H)).
Thus, the limit $${1\over N^n}\nu_\nn^\ND$$ exists and is called the density
of N-D eigenvalues and we denote it  by $\mu^\ND$.

\subsection{Statements and proofs of the results.}
We state 3 elementary results on the spectrum of these
operators and their relations with the density of states and the 
density of N-D eigenvalues.
The first one is the counterpart of a classical result for random
Schr\"odinger operators (but nevertheless never appeared in the
litterature).
For convenience, we suppose here
the existence of the density of states.
\ali
We denote by $\Sigma^\pm$ the topological 
spectrum of the operators $H_\infi^\pm$
(and we simply write $\Sigma$ when $\partial F_\infi=\emptyset$).
 We recall that the essential spectrum is
obtained from the spectrum by removing all isolated points corresponding
to eigenvalues with finite multiplicity, we denote it by $\Sigma^\pm_{ess}$.
\begin{propos}
\label{p.1}
For both the discrete and the continuous case we have the following:

i) If the boundary set $\partial X_\infi=\partial F_\infi$ is
empty  then
$\supp \mu = \Sigma=\Sigma_{ess}$.

ii) Otherwise we just have
$\supp \mu =\Sigma^+_{\hbox{ess}}=\Sigma^-_{\hbox{ess}}$.
Moreover, the eigenvalues eventually lying in
$\Sigma^\pm\setminus \supp (\mu)$ have multiplicity 1.
\end{propos}
\Rm: We are in case i) for almost all
blow-up $\w$, for the product of the uniform measure on $\unN$.
\ali
\Rm:
In \cite{Sabot4} we proved i) in the case of Nested fractals for equal weights
$\alpha_i=\alpha$ and $\beta_i={1\over N}$.
For this class of self-similar sets, due to symmetry arguments, we proved
that the equality
$\supp \mu =\Sigma^+=\Sigma^-$ is true even in the case ii).
In \cite{Sabot7}, we plan to show that in the case of the unit interval
blowed-up to the half-line $\BR_+$ (by the constant blow-up $\w_k=1$)
the spectrum of the operator can be pure point with isolated
eigenvalues of multiplicity  1 lying in the complement of
$\supp \mu$ and accumulating on $\supp \mu$. Therefore in this case the equality
$\Sigma_{\hbox{ess}}^\pm=\supp \mu$ is satisfied by not $\Sigma^\pm=\supp \mu$.
\ali
Proof:
The proof is similar to that of \cite{Sabot4}, except that
we must be carreful with the inhomogeneous weights $\alpha_i$, $\beta_i$
and that we must prove the extra result ii).
We consider first the discrete case.
We know from section 1.2 that the norm of the operator
$ H_\infi^\pm$ in
$L^2(F_\infi, b_\infi)$ is finite, say smaller than a real $K>0$.
\ali
By classical arguments we know that $\supp (\mu)\subset \Sigma_{ess}^\pm$.
We denote by $P_\infi^\pm(d\lambda)$ and $P^\pm_\nn (d\lambda)$ 
the spectral resolution of the
operators $H^\pm_\infi$ and $H_\nn^\pm$ resp. on
$\ddd_\infi^\pm$ and $\ddd_\nn^\pm$.
We first prove i).
We suppose that $\partial F_\infi =\emptyset$ and we let
$\lambda\in \Sigma$ and $\epsilon>0$.
We choose $f$ in $P_\infi([\lambda-\epsilon/4 ,\lambda+\epsilon/4])(\ddd_\infi)$
such that $\int f^2 db_\infi =1$.
For all $\eta>0$ we can find $n_0$ such that
$\int_{F_\infi\setminus \intF_{<n_0>}} \vert f\vert^2 db_\infi \le \eta$.
We define $\tilde f$ by $\tilde f=f$ on $\intF_{<n_0>}$ and $\tilde f=0$ on
$F_\infi\setminus \intF_{<n_0>}$.
We have easily
$$\int \vert H_\infi \tilde f-\lambda\tilde f\vert^2 db_\infi\le
K^2\eta +\epsilon^2/16+\lambda^2\eta,
$$
and $\int \vert \tilde f\vert^2 db_\infi \ge 1-\eta$.
\ali
Set $\hat f={\tilde f\over (\int \vert \tilde f\vert^2 db_\infi )^\demi}$.
For $p_0$ large enough $\hat f$
is in $\ddd_{H_{<n_0+p_0>}^+}\cap \ddd_{H_{<n_0+p_0>}^-}$
(precisely, it is sufficient 
that $\partial F_{<n_0+p_0>}\cap \partial F_{<n_0>}=\emptyset$,
which is possible since $\partial F_\infi=\emptyset$).
We can choose $\eta$ such that
\beq
\label{f.e.1}
\int \vert H_\infi \hat f-\lambda \hat f\vert^2 db_\infi\le \epsilon^2/4.
\eeq
Then we proceed as in lemma 2.1. of \cite{Sabot4}.
Precisely, equation (\ref{f.e.1}) and the fact that $\hat f$ is in
$\ddd_{H_{<n_0+p_0>}^+}\cap \ddd_{H_{<n_0+p_0>}^-}$, imply
that
$$\| P^\pm_{<n_0+p_0>}([\lambda-\epsilon,\lambda+\epsilon])(\hat f)\|\ge {1\over 4}.
$$
At each level $<n_0+p_0+k>$ we can make $N^k$ copies of $\hat f$
on the
$<n_0+p_0>$-cells of
$F_{<n_0+p_0+k>}$.
This implies that
$$\int_{[{\lambda-\epsilon},{\lambda+\epsilon}]} \nu_{<n_0+p_0+k>}^\pm
=\hbox{Tr}(P_{<n_0+p_0+k>}^\pm([\lambda-\epsilon,\lambda+\epsilon]))
\ge {1\over 4} N^k.
$$
This implies that $\mu([\lambda-\epsilon,\lambda+\epsilon])\ge {1\over 4N^{n_0+p_0}}>0$.
Hence, we proved that $\lambda\in \supp(\mu)$.
\ali
To prove ii) it is enough to prove that if $\lambda\in \Sigma$ is such that
for all $\epsilon>0$ $\dim P^\pm_\infi([\lambda-\epsilon ,\lambda+\epsilon])
(\ddd_\infi^\pm)\ge 2$
then $\lambda$ is in $\supp \mu$.
We do the proof for the Neuman boudary condition, 
the proof for the Dirichlet boundary
condition being identical.
If $\partial F_\infi \ne \emptyset$ then
$\partial F_\infi$ contains a unique point
that we denote $z_0$.
Let $\epsilon>0$ and suppose that
$\dim P^+_\infi([\lambda-\epsilon ,\lambda+\epsilon])(\ddd_\infi^+)\ge 2$.
We can find $f$, with $L^2$ norm 1,
in $P^+_\infi([\lambda-\epsilon/4 ,\lambda+\epsilon/4])(\ddd_\infi^+)$
such that $f(z_0)=0$.
Then, exactely as previously we can construct $\hat f$ with norm 1, proportional
 to
$f$ on $F_{<n_0>}$, null outside, and  such that
$\int \vert H_\infi^+ \hat f-\lambda \hat f\vert^2 db_\infi\le \epsilon^2/4$.
Moreover, we see that $\hat f$ is in $\ddd_{H_{<n_0+1>}^+}\cap \ddd_{H_{<n_0+1>}
^-}$
(indeed, $\partial F_{<n_0+1>}\cap \partial F_{<n_0>}=\{z_0\}$ and
$\hat f(z_0)=0$).
At this point the proof goes exactly as before.
\ali
The proof in the continuous case is 
more technical since we cannot just approximate
the function $f$ by its restriction to $X_\nn$ 
(which is in general not in the domain
of $H_\infi$), but we need to approximate it smoothly as it is done
in \cite{Sabot4}. We safely leave the details to the reader since the
proof given for the discrete case for ii) and the 
arguments developped in \cite{Sabot4} give easily the result. $\square$
\ali

The following results concern the Lebesgue decomposition of the
spectrum.
Let us first recall some elementary notions.
We denote by $P_\infi^\pm(d\lambda)$ the spectral resolution of the
self-adjoint operator $H^\pm_\infi$ on the Hilbert space $\ddd_\infi^\pm$.
If $f$ is in $\ddd_\infi^\pm$, remind that the spectral measure
of $f$ is defined as the measure $\sigma^\pm(f)(d\lambda)$ on
$\BR$ given by
$$\sigma^\pm(f)(A) =\|P_\infi^\pm(A)(f)\|^2,$$
for any borelian $A\subset \BR$, where $\| \|$ denotes the
$L^2$ scalar product associated with $b_\infi$ on $F_\infi$ in the discrete
setting, and $m_\infi$ on $X_\infi$ in the continuous setting.
We denote by $\sigma^\pm_{ac}(f)(d\lambda)$, $\sigma^\pm_{sc}(f)(d\lambda)$,
$\sigma^\pm_{pp}(f) (d\lambda)$ respectively the absolutely continuous,
the singular continuous, the purely ponctual part of the Lebesgue
decomposition of the measure $\sigma^\pm(f)(d\lambda)$.
Remind that the Hilbert space $\ddd_\infi^\pm$ can be decomposed
into three orthogonal Hilbert subspaces (cf for example \cite{CarmonaL})
$$\ddd_\infi^\pm=\hhh_{ac}^\pm\oplus \hhh^\pm_{sc}\oplus \hhh^\pm_{pp},
$$
such that $f$ is in $\hhh_{ac}^\pm$, $\hhh^\pm_{sc}$, $\hhh^\pm_{pp}$
iff its spectral measure is respectively
absolutely continuous, singular continuous or  purely ponctual.
The Lebesgue decomposition of the spectrum is the closed sets
$\Sigma_{ac}^\pm$, $\Sigma_{sc}^\pm$, $\Sigma_{pp}^\pm$
equal to the topological spectrum of the restriction of
$H_\infi^\pm$ to the subspaces $\hhh_{ac}^\pm$, $\hhh^\pm_{sc}$, $\hhh^\pm_{pp}$.
It is clear that the subspaces $\hhh_{\ND}$ generated by
the Neuman-Dirichlet eigenfunctions is included in both
$\hhh_{pp}^-$ and $\hhh_{pp}^+$.
It is then natural to define $\tilde \hhh_{pp}^\pm$ as the orthogonal
supplement of $\hhh_{ND}$ in
$\hhh_{pp}^\pm$ and to define $\tilde \Sigma_{pp}^\pm$  and
$\Sigma_\ND$ as
the topological spectrum of $H_\infi^\pm$ restricted
respectively to $\tilde \hhh_{pp}$ and $\hhh_{ND}$.
It is clear by definition that $\Sigma_\ND=\supp \mu^\ND$
(and this is true for any blow-up $\w$).

As pointed out, the infinite lattices $F_\infi$
(or the unbounded set $X_\infi$) are not isomorphic for different 
blow-up $\w$.
Hence, the spectral properties of $H_\infi^\pm$ depends a priori on
$\w$: to show this dependence we sometimes write
$P_\infi^\pm(\w,d\lambda)$, $\Sigma^\pm_\cdot(\w)$, $\cdots$.

We endow $\W=\{1,\ldots ,N\}^\BN$ with the product of the
uniform measure on $\{1,\ldots ,N\}$.
In the next two propositions we give almost sure results on the blow-up.
The following result is the analogous of a result
initially due to Pastur, \cite{Pastur1}, for random Schr\"odinger operators.
\begin{propos}
\label{p.2}
There exist deterministic sets $\Sigma$, $\Sigma_{ac}$, $\Sigma_{sc}$,
$\Sigma_{pp}$, $\tilde \Sigma_{pp}$ and $\Sigma_{\ND}$
 such that for almost all $\w$ in $\W$
(for the product of the uniform measure on $\unN$)
we have 
$$
\Sigma^\pm(\w)=\Sigma,\;\; \Sigma_\cdot^\pm (\w)=\Sigma_\cdot,
\;\; \tilde \Sigma_{pp}^\pm(\w)=\tilde \Sigma_{pp}.$$
\end{propos}
\Rm:
\label{r.2}
As we pointed out, $\Sigma_{\ND}$ is constant in $\w$ and equal to
$\supp(\mu^\ND)$.
From proposition \ref{p.1} we see that the almost sure
spectrum $\Sigma$ is equal to $\supp \mu$.
\ali
\Rm:
The structure of the spectrum can really depend on $\w$.
In a forthcoming paper, \cite{Sabot7}, we plan to prove that 
for a self-similar Sturm-Liouville operator on $[0,1]$, the spectrum
is continuous for a typical blow-up $\w$, but can be pure point for
a particular $\w$.
\ali
Proof:
Remark first that for almost all blow-up $\w$, $\partial F_\infi=\emptyset$.
It is thus enough to consider only $\w$ such that $\partial F_\infi=\emptyset$.
Let $P_{\infi}(d\lambda,\w)$ denotes the spectral resolution of $H_\infi(\w)$.
Denote by $\tilde \hhh$ the orthogonal supplement 
of $\hhh_\ND$ in
$\ddd_\infi$.
For a function $f$ in $\ddd_\infi$ we denote by
$\tilde \sigma (f)(d\lambda, \w)$ the spectral measure of the projection
of $f$ on $\tilde \hhh$.
We denote by $\tilde \sigma_{ac}(f)(d\lambda,\w)$, $\tilde \sigma_{sc}(f)(d\lambda,\w)$,
$\tilde\sigma_{pp}(f)(d\lambda,\w)$, 
resp. the absolutely continuous, the singular continuous
and the purely ponctual part of the Lebesgue decomposition of the measure
$\tilde \sigma(f)$.
Remark that $\tilde \sigma_{ac}(f)=\sigma_{ac}(f)$
and $\tilde \sigma_{sc}(f)=\sigma_{sc}(f)$
($\sigma_{pp}(f)$ is the sum of $\tilde \sigma_{pp}(f)$
plus the spectral measure of the projection of $f$ on $\hhh_\ND$).
Consider $x$ in  $F_\infi$, we first prove that  the map
$\w\rightarrow \tilde \sigma(\delta_x)(d\lambda,\w)$,
where $\delta_x$ is the Dirac function at $x$, is measurable in $\w$
(the $\sigma$-field on the set of non-negative measures on $\BR$ is the
smalest $\sigma$-field such that $\mu \rightarrow \mu(A)$ is measurable
for any Borelian $A$).
Indeed, denote by $\tilde E_\nn^+$ the orthogonal supplement
(for the $L^2$ scalar product for $b_\nn$) of $E^\ND_\nn$ in
$\ddd_\nn^+$, and for $x$ in $F_\nn$ and $m\ge n$, by
$\tilde \sigma^+_{<m>}(\delta_x)(d\lambda,\w)$ the spectral measure of
the projection of $\delta_x$ on $\tilde E^+_{<m>}$ for the operator
$H_{<m>}^+$. It is clear that $\tilde\sigma(\delta_x)(d\lambda,\w)$ is the
limit of $\tilde \sigma^+_{<m>}(\delta_x)(d\lambda,\w)$ when $m$ goes to infinity.
But $\tilde \sigma^+_{<m>}(\delta_x)(d\lambda,\w)$ depends on $\w$ only by
the finite sequence $(\w_1,\ldots ,\w_m)$ (which gives the position of
$x\in F_\nn$ as a point in $F_{<m>}$),
and hence, is
measurable in $\w$. This implies the measurability of
$\w\rightarrow \tilde \sigma(\delta_x)(d\lambda,\w)$.
Proceeding exactely like in lemma V.15 of \cite{CarmonaL}, we know that
the components $\tilde \sigma_{ac}(\delta_x)(d\lambda,\w)$,
$\tilde \sigma_{sc}(\delta_x)(d\lambda,\w)$, 
$\tilde \sigma_{pp}(\delta_x)(d\lambda,\w)$  
of the Lebesgue decomposition of
$\tilde \sigma(\delta_x)(d\lambda,\w)$ are measurable.
\ali
Consider now the familly of "translations" in $\W$ defined by
$$\tau_{i_1,\ldots ,i_k}(\w)=(\w_1+i_1[N],\ldots ,\w_k+i_k[N], \w_{k+1},\ldots ),$$
where $\w_j+i_j[N]$ is the value of $\w_j+i_j$ modulo $N$.
It is clear that 
$\{\tau_{i_1,\ldots ,i_k}\}_{k\in \BN\atop i_1,\ldots ,i_k\in \unN^k}$
is a  measure preserving ergodic familly of transformations of $\W$.
Writting
\beqn
\{\w\;\hbox{ s.t. } \; \lambda \in \Sigma_\cdot (\w)\}=
\cap_{\lambda',\lambda''\in \BQ\atop \lambda'<\lambda<\lambda''}
\{\w\;\hbox{ s.t. } \; \exists n, \;\exists x\in F_\nn,\;
\sigma_\cdot (\delta_x)(]\lambda',\lambda''[,\w)>0\},
\eeqn
(and idem for $\tilde \Sigma_{pp}(\w)$)
we know that the set 
$\{\w\; \hbox{ s.t. } \; \lambda \in \Sigma_\cdot (\w)\}$ 
is measurable and  $\tau$
invariant, and thus is of measure 0 or 1.
We define the deterministic set
$\Sigma_\cdot =\overline 
{\{\lambda\in \BQ\; \hbox{ s.t. } \; \BP(\lambda \in \Sigma_\cdot (\w))=1\}}$,
where $\BP$ denotes the expectation with respect to $\w$.
It is clear, since $\Sigma_\cdot (\w)$ 
is closed, that we have $\Sigma_\cdot (\w)=\Sigma_\cdot$
for allmost all $\w$. $\square$

\begin{propos}
\label{p.3}
If the density of states is completely created by the N-D eigenvalues, i.e. if
$\mu^\ND =\mu$ then for almost all $\w$ in $\W$
the set of N-D is
complete i.e. $\hhh_\ND = \ddd_\infi^+(\w)=\ddd_\infi^-(\w)$.
In this case, 
the spectrum of $H_{\infi}$ is then pure point with compactly supported eigenfunctions.
\end{propos}
\Rm: When this last equality is satisfied then necessarily
$\partial X_\infi=\partial F_\infi=\emptyset$.
In particular, this means that if $\partial F_{\infi}\ne \emptyset$
then there is a component of the spectrum in the complement of
$\hhh_\ND$ (for the Sierpinski gasket, it is known from \cite{Teplyaev}
that for Neuman boundary condition, this component is also pure
point, but the answer is not known for the Dirichlet boundary condition).
\ali
\Rm: It is known that the equality $\mu^\ND =\mu$ is satisfied for
Nested fractals (cf \cite{Sabot4}, \cite{Sabot5}).
In \cite{Sabot4} we precised an almost sure  class of blow-ups for which the
set of N-D eigenfunctions is complete (called asymmetrical blow-ups). 
In particular for
the Sierpinski gasket this is true as soon as $\partial F_\infi=\emptyset$
(but this was known since \cite{Teplyaev}).
\ali
Proof:
Denote by $\BP$ and $\BE$ the probability and the expectation with
respect to the blow-up for the product of the uniform measure
on $\unN$. We first prove the result
in the discrete case.
We must prove that
\beqn
&&
\BP (\overline{\cup E_\nn^\ND} =\ddd^+_\infi )=1,
\\
&\Leftrightarrow &
\BP(\forall f\in \ddd_\infi \hbox{ with compact support}, \;\;
\lim_{n\to\infty} \| P_{\tilde E^+_\nn} f\|_{b_\nn}^2 = 0)=1,
\eeqn
where $P_{\tilde E^+_\nn}$ is the orthogonal projection on the space $\tilde E_\nn^+$
generated by the "Neumann only" eigenfunctions, i.e. the orthogonal
supplement in $\ddd_\nn^+$ of $E_\nn^\ND$.
But this is again equivalent to
\beqn
&& \forall k_0, \; \forall f\in \ddd_{<k_0>}^-,\;\; 
\BP(\lim_{n\to\infty} \|P_{\tilde E^+_\nn} f\|^2_{b_\nn} = 0)=1
\\
&\Leftrightarrow &
\forall k_0, \; \forall f\in \ddd_{<k_0>}^-,\;\forall \epsilon >0,
\;\; \lim_{n\to\infty} \BP(\|P_{\tilde E^+_\nn} f\|^2_{b_\nn} \ge \epsilon ) = 0.
\eeqn
Let us now make a remark:
for different blow-ups $\w$, the sets $F_\nn$ are isomorphic but
the measures $b_\nn$ are not equal but differ from a constant
multiple.
On $F_\nn$, we introduce the measure $\tilde b_\nn$, independant of
$\w$, by
$$\int_{F_\nn} f d\tilde b_\nn=
\sum_{i_1,\ldots ,i_n} \beta_{i_1}\cdots \beta_{i_n} \int_F f_{|F_{\nn,i_1,\ldots ,i_n}} db.
$$
Remark that $b_\nn=\beta_{\w_1}^{-1}\cdots \beta_{\w_n}^{-1} \tilde b_\nn$.
Choose now a basis $g_1,\ldots , g_{\dim \tilde E^+_\nn}$ of $\tilde E^+_\nn$,
orthonormal for the scalar product associated with $\tilde b_\nn$.

We have 
\beqn
\BP(\|P_{\tilde E^+_\nn} f\|_{b_\nn}^2\ge \epsilon )
&\le& {1\over \epsilon}
\BE( \|P_{\tilde E^+_\nn} f\|_{b_\nn}^2)          
\\
&=&
\BE ( \beta_{\w_1}^{-1}\cdots \beta_{\w_n}^{-1} \|P_{\tilde E^+_\nn} f\|^2_{\tilde b_\nn})
\\
&=&
\sum_{i=1}^{\dim \tilde E^+_\nn}
\BE( \beta_{\w_1}^{-1}\cdots \beta_{\w_n}^{-1} \vert <f, g_i>_{\tilde b_\nn}\vert^2).
\eeqn
Now, we average on the blow-up: to do this we consider that
the cell $F_{<k_0>}$  is the subcell $F_{\nn, w_n, \ldots , w_{k_0+1}}$
of $F_{\nn}$ and we average on the position of $F_{<k_0>}$.
We denote by $f_{w_n, \ldots ,w_{k_0+1}}$ the function with value
$f$ on the cell $F_{\nn, w_n,\ldots ,w_{k_0+1}}$ and 0 outside.
The last expression is equal to
\beqn
{1\over N^n}
\sum_{\w_1,\ldots ,\w_{k_0}} \beta_{\w_1}^{-1}\cdots \beta_{\w_{k_0}}^{-1}
\sum_{\w_{k_0+1},\ldots ,\w_n} \sum_{i=1}^{\dim\tilde E_\nn^+}
b_{\w_{k_0+1}}^{-1}\cdots b_{\w_n}^{-1}
<f_{\w_{k_0+1},\ldots ,\w_n},g_i>_{\tilde b_\nn}.
\eeqn
But the $f_{\w_{k_0},\ldots ,\w_n}$ are orthogonal and
$$\|f_{\w_{k_0},\ldots ,\w_n}\|_{\tilde b_\nn}^2=
\beta_{\w_{k_0+1}}^{-1}\cdots \beta_{\w_n}^{-1} \|f\|_{\tilde b_{<k_0>}}^2.
$$
This implies that
\beqn
\sum_{\w_{k_0+1},\ldots ,\w_n}
\beta_{\w_{k_0}}^{-1}\cdots \beta_{\w_n}^{-1}
\vert <f_{\w_{k_0+1},\ldots ,\w_n},g_i>_{\tilde b_\nn}\vert^2
\le \|f\|^2_{\tilde b_{<k_0>}},
\eeqn
for all $i$,
and thus that
\beqn
\BE( \|P_{\tilde E^+_\nn} f\|_{b_\nn}^2)
&\le&
{\dim \tilde E^+_\nn\over N^{n-k_0}
}
{1\over N^{k_0}}
\sum_{\w_1,\ldots ,\w_{k_0}} 
\beta_{\w_1}^{-1}\cdots \beta_{\w_{k_0}}^{-1}
\|f\|_{\tilde b_{<k_0>}}^2
\\
&=&
{\dim \tilde E^+_\nn\over N^{n-k_0}
}
\BE (\|f\|^2_{b_{<k_0>}}).
\eeqn
which goes to 0 when $n$ goes to infinity, by hypothesis.
Thus, we proved that almost surely on the blow-up
we have $\overline{\cup  E^\ND_\nn }=\ddd^+_\infi$. When this
situation is satisfied then necessarily $\ddd_\infi^+=\ddd_\infi^-$ and
hence $\partial F_\infi =\emptyset$
(since
$\overline{\cup  E^\ND_\nn }\subset \ddd_\infi^-$).
Moreover the spectrum of $H_\infi$ is pure point since the Neumann-Dirichlet
eigenfunctions form a dense set of compactly supported eigenfunctions.

To prove the result in the continuous case we must consider the space
$\tilde E_\nn^+(\lambda)$, $\lambda\le 0$, generated by the "Neuman only" 
eigenfunctions with
eigenvalues larger than $\lambda$, i.e. the orthogonal
supplement in $P_\nn^+([\lambda, 0]) (\ddd_\nn^+)$ of
$E_\nn^\ND \cap P_\nn^+([\lambda, 0]) (\ddd_\nn^+)$.
The space $\tilde E_\nn^+(\lambda)$ is finite dimensionnal and we
prove exactly in the same way that almost surely on the blow-up
for any function $f$ with compact support 
$\|P_{\tilde E_\nn^+(\lambda)}f\|^2$ converges
to 0. $\square$

\subsection{About a "Lebesque decomposition" of the density of states}
\subsubsection{The discrete case}
To avoid confusion, we precise that  we do not consider here the Lebesgue
decomposition of the measure $\mu$, but a decomposition of the measure
$\mu$ into three (or four) parts corresponding to the Lebesgue
decomposition of the spectral measures.
As pointed out in remark \ref{r.2}, we proved that the measures
$\mu$ and $\mu^\ND$ are related to the almost sure spectrum and
to the Neuman-Dirichlet spectrum by 
$\supp(\mu)=\Sigma$, $\Sigma_{\ND}=\supp \mu^\ND$.
In \cite{Sabot5}, we were able to compute these two measures
in terms of a certain renormalization map that we explicitely defined.
The aim of this section is to introduce some measures generalizing
the measures $\mu$ and $\mu^\ND$ to the different parts of
the Lebesgue decomposition of the spectrum.
We are not able to say much about these measures, but we think
they are central notions. In particular, we think that the central
question is wether it is possible to compute these measures in terms
of the renormalization map introduced in \cite{Sabot5}.

Let us start by a lemma.
\begin{lem}
\label{l.1}
The measures $\mu$ and $\mu^\ND$ satisfy:
\beq
\mu(d\lambda)=
\BE\left(
\sum_{x \in F_{<0>}}
{b_{<0>}(x)\over (b_\infi(x))^2}
\sigma(\delta_x)(d\lambda, \w)
\right),
\\
\mu^\ND(d\lambda)=
\BE\left(
\sum_{x \in F_{<0>}}
{b_{<0>}(x)\over (b_\infi(x))^2}
\sigma_{\ND} (\delta_x)(d\lambda, \w)
\right),
\eeq
where $\BE$ denotes the expectation with respect to the 
blow-up $\w\in\unN^\BN$,
and $\sigma(\delta_x)$ and $\sigma_\ND(\delta_x)$ denotes respectively
the spectral measure of the Dirac function $\delta_x$ at $x$ and the spectral
measure of the projection of $\delta_x$ to the subspace $\hhh_\ND$ generated by
the Neuman-Dirichlet eigenfunctions.
\end{lem}
\noindent
N.B.: Remind that for almost all $\w$, $\partial F_\infi=\emptyset$ so that
the boundary condition does not matter when we take expectation and we can simply
write $\sigma(\delta_x)(d\lambda, \w)$ for $\sigma^\pm(\delta_x)(d\lambda, \w)$.
\ali
Proof:
We first present the proof for $\mu$.
Remind that $\sigma^+_\nn(\delta_x)(d\lambda,\w)$ is the spectral measure 
of the Dirac function $\delta_x$ for the operator
$H_\nn^+$ on $\ddd_\nn^+$ (it depends on $(\w_1,\ldots ,\w_n)$
via the position of $F_{<0>}$ in $F_\nn$).
It is clear that for all $\w$ such that $\partial F_\infi=\emptyset$
and all $x$ in $F_{<0>}$
we have
$${b_{<0>}(x)\over (b_\infi(x))^2}
\sigma(\delta_x)(d\lambda, \w)=\lim_{n\to\infty} 
{b_{<0>}(x)\over (b_\nn(x))^2}
\sigma_\nn^+(\delta_x)(d\lambda, \w).
$$
Choose now a basis of eigenfunctions 
of $H_\nn^+$,
$g_1^+,\ldots ,g_{\vert F_\nn\vert }^+$
for the eigenvalues $\lambda^+_1=0>\cdots \ge \lambda^+_{\vert F_\nn\vert }$.
Assume moreover that these eigenfunctions are orthonormal for the
$l^2$ scalar product associated with the measure $\tilde b_\nn$ on $F_\nn$,
introduced in the proof of proposition \ref{p.3} 
(and remind that $b_\nn=\beta_{\w_1}^{-1}\cdots \beta_{\w_n}^{-1}\tilde b_\nn$).
Each point in $F_\nn$ can be labelled in a natural way
(and in a non unique way)
by a point $(\w_1,\ldots ,\w_n,z)$ in $\unN\times F$.
In the following, by abuse of notations, 
we simply write $(\w_1,\ldots ,\w_n,z)$
for the corresponding point in $F_\nn$.
We have
\beqn
&&
\BE\left(
\sum_{x \in F_{<0>}}
{b_{<0>}(x)\over (b_\nn(x))^2}
\sigma^+_\nn (\delta_x)(d\lambda, \w)
\right)
\\
&=&
\BE
\left(
\sum_{x \in F_{<0>}}
{b_{<0>}(x)\over (b_\nn(x))^2}
\sum_{i=1}^{\vert F_\nn\vert}
{\vert <g_i^+, \delta_x>_{b_\nn}\vert^2
\over \| g_i\|_{b_\nn}^2} \delta_{\lambda_i^+}(d\lambda)
\right)
\\
&=&
\BE
\left(
\sum_{x \in F_{<0>}}
\sum_{i=1}^{\vert F_\nn\vert}
b_{<0>}(x)
\beta_{\w_1}\cdots \beta_{\w_n} \vert g_i^+(x)\vert^2 \delta_{\lambda_i^+}(d\lambda)
\right)
\\
&=&
{1\over N^n}\sum_{i=1}^{\vert F_\nn\vert}
\sum_{\w_1,\ldots ,\w_n=1}^N\sum_{z\in F}
\beta_{\w_1}\cdots \beta_{\w_n} b(z) 
\vert g_i^+((\w_1,\ldots ,\w_n,z)\vert^2 \delta_{\lambda_i^+}(d\lambda)
\\
&=&
{1\over N^n}
\sum_{i=1}^{\vert F_\nn\vert}
\delta_{\lambda_i^+}(d\lambda)
\\
&=&
{1\over N^n}
\nu^+_\nn(d\lambda).
\eeqn
This proves the formula concerning $\mu$.
The proof for the formula for $\mu^\ND$ is similar: on just has to replace
$\sigma_\nn^+(\delta)$ by $\sigma_\nn^\ND (\delta_x)$, the spectral measure
of the projection of $\delta_x$ on $E_\nn^\ND$.
$\square$
\begin{defin}
\label{d.1}
We introduce the measures $\mu^{ac}$, $\mu^{sc}$, $\mu^{pp}$, $\tilde \mu^{pp}$,
by
\beq
\mu^\cdot
(d\lambda)=
\BE\left(
\sum_{x \in F_{<0>}}
{b_{<0>}(x)\over (b_\infi(x))^2}
\sigma_\cdot
(\delta_x)(d\lambda, \w)
\right),
\eeq
and idem for $\tilde \mu^{pp}$.
Remark that we have
\beqn
&\mu=\mu^{ac}+\mu^{sc}+\mu^{pp},&
\\
&\mu^{pp}=\tilde \mu^{pp}+\mu^\ND.&
\eeqn
\end{defin}
Then we have the following easy proposition:
\begin{propos}
\label{p.4}
We have
\beqn
& \Sigma_\cdot =\supp (\mu^\cdot ),&
\\
&\tilde \Sigma_{pp}=\supp(\tilde \mu^{pp}),&
\eeqn
where $\Sigma_{ac}, \Sigma_{sc},\Sigma_{pp}, \tilde \Sigma_{pp}$ are
the almost sure components of the spectrum introduced in
proposition \ref{p.2}.
\end{propos}
Proof:
By the definition itself it is clear that
$\Sigma_\cdot\subset \supp(\mu_\cdot)$.
Reciproquely, if $\Sigma_\cdot\cap ]\lambda',\lambda''[\ne \emptyset$
this means that there exist $n$ and $x$ in $F_\nn$ such that
$\sigma_\cdot (\delta_x)(]\lambda',\lambda''[,\w)>0$ for a set
of blow-up $\w$ of positive measure.
But the point $x$ in $F_\nn$ is in $F_{<0>}$ for all $\w$ starting
from a certain sequence $(\w_1,\ldots ,\w_n)$.
This means that for a set of positive measure 
of blow-up $\w$, there exists $x$ in $F_{<0>}$
such that $\sigma_\cdot (\delta_x)(]\lambda',\lambda''[,\w)>0$.
This immediately implies that $\mu^\cdot (]\lambda',\lambda''[)>0$
and thus that $\supp \mu^\cdot =\Sigma_\cdot$.
$\square$
\ali

In \cite{Sabot5} we were able to compute the measures $\mu$ and
$\mu^\ND$ and to caracterize the equality $\mu=\mu^\ND$.
The natural question (but certainly difficult) is wether it is
possible to compute the different measures $\mu^{ac}$, $\mu^{sc}$,
$\mu^{pp}$, $\tilde \mu^{pp}$ in terms of the renormalization
map we introduced in \cite{Sabot5}.
This would give an information stronger than the Lebesgue decomposition
of the spectrum.
In particular it would be interesting to understand the measure
$\tilde \mu^{pp}$, or the set $\tilde \Sigma^{pp}$, which corresponds
to the pure point part not induced by the N-D spectrum.
There are very few examples where the spectral properties of the operator $H_\infi$
are understood.
There is the case where $\mu^\ND=\mu$ (including the Sierpinski gasket
and the  nested fractals, cf \cite{Teplyaev}, \cite{Sabot4}),
corresponding more or less to the case where the asymptotic degree
$d_\infty$ is smaller than $N$ (cf \cite{Sabot5}).
In this case we know that the only non-empty component in the spectrum is
the Neuman-Dirichlet component.
We can prove also that in the case of a self-similar Sturm-Liouville
operator, the spectrum is continuous, i.e. $\Sigma_{pp}=\emptyset$.
We do not know any example where $\tilde \Sigma_{pp}$ is non empty.

Of course, what make things easy for the measure $\mu$ and $\mu^\ND$
is that they have an expression as a limit of counting measures
related to the operators on finite level $F_\nn$.
There is no corresponding expression for the other parts of the 
density of states $\mu^{ac}$, $\mu^{sc}$, $\tilde \mu^{pp}$.

\subsubsection{The continuous case}
Of course, there are similar formulas for the continous setting.
We give without proof the correct definitions for the measures
$\mu^{ac}$, $\mu^{sc}$, $\mu^{pp}$, $\tilde \mu^{pp}$.
Denote by $R_{X_{<0>}}: L^2(X_\infi)\rightarrow L^2(X_\infi)$ 
the operator of restriction to $X_{<0>}$ defined by
$R_{X_{<0>}}(f)(x)={\indic}_{\{x\in X_{<0>}\}}f(x)$.
We also denote respectively by
$P_\infi^{ac}(d\lambda,\w)$,
$P_\infi^{sc}(d\lambda,\w)$,
$P_\infi^{pp}(d\lambda,\w)$,
$P_\infi^{\ND}(d\lambda,\w)$,
$\tilde P_\infi^{pp}(d\lambda,\w)$,
the composition of the spectral resolution $P_\infi(d\lambda,\w)$
with the projection on respectively
$\hhh_{ac}$, $\hhh_{sc}$, $\hhh_{pp}$, $\hhh_\ND$,
$\tilde \hhh_{pp}$.
Then we define
$$\mu^\cdot=
\BE\left(
\hbox{Trace} ( R_{X_{<0>}}\circ P^\cdot_\infi(d\lambda,\w)\circ R_{X_{<0>}}
)\right),
$$
and idem for $\tilde \mu^{pp}$.
It is not difficult to prove that these measures have the same  properties
as in the discrete case (i.e. the analogous of lemma \ref{l.1} and proposition
\ref{p.4} are satified).
\ali\ali
\ali
{\it Acknowledgment:}
I thank Fran\c cois Ledrappier who suggested me to
consider random blow-ups.

\end{document}